\documentclass[aps,prc,twocolumn,amsfonts,amsmath,amssymb,showpacs,floatfix]{revtex4-1}

\usepackage{graphicx}
\usepackage{longtable}
\usepackage{bm}
\usepackage{multirow}
\usepackage{hyperref}
\usepackage{ulem}
\hypersetup{colorlinks,citecolor=blue,filecolor=blue,linkcolor=blue,urlcolor=blue}
\usepackage{color}
\usepackage{dcolumn}

\begin{document}

\title{Neutron {\boldmath$s$} States in Loosely Bound Nuclei}

\author{C.~R.~Hoffman}
\affiliation{Physics Division, Argonne National Laboratory, Argonne, Illinois 60439, USA}
\author{B.~P.~Kay}
\affiliation{Physics Division, Argonne National Laboratory, Argonne, Illinois 60439, USA}
\author{J.~P.~Schiffer}
\affiliation{Physics Division, Argonne National Laboratory, Argonne, Illinois 60439, USA}

%\date{\today}

\begin{abstract}

In reviewing the data that has accumulated in light nuclei we find that the binding energy plays a critical role in describing the variation in energy of $s$ states relative to other states. The behavior of states with zero angular momentum within a few MeV of threshold is qualitatively different from that of neutron states with any other $\ell$ value or of any proton state. This observation is explored for simple Woods-Saxon potentials and is remarkably successful in describing a wealth of experimental data for nuclei with neutron numbers between 5 and 10. The lingering of neutron $s$ states just below threshold is associated with the increases in radii of the neutron density distributions, the neutron halos, and leads to speculations about possible halos in heavier nuclei.

\end{abstract}


\maketitle

A description for the evolution of nuclear excitations with neutron excess, from tightly bound stable systems to loosely bound exotic ones, is a major challenge to our understanding of nuclear structure. Startling modifications to the spacing and sequence of single-particle excitations are now experimentally well documented~\cite{sorlin} and closed shells with `magic numbers' of nucleons, once thought to apply to all nuclei, are now known to change when moving away from stability.

\begin{figure}
\centering
\includegraphics[scale=0.35]{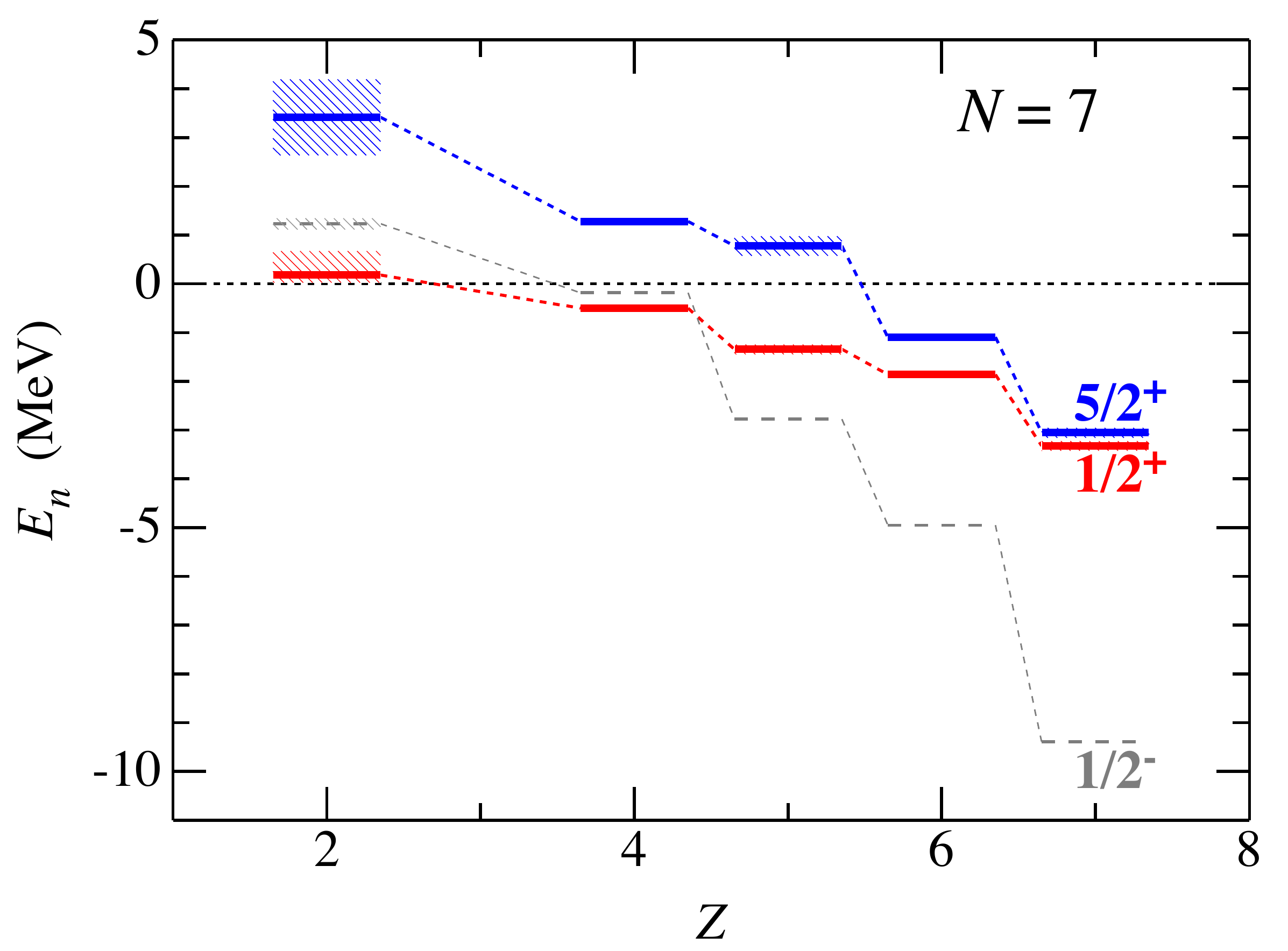}
\caption{\label{fig1}(color online). The experimental data available on the energy $E_{n}$, relative to the neutron threshold, of the $0p_{1/2}$, $1s_{1/2}$, and $0d_{5/2}$ states in $N=7$ nuclei. The sources of the data are given in the Appendix. Uncertainties are indicated by shading.}
\end{figure}

In light stable nuclei the $0p$-shell closes with eight nucleons, accounting for the stability of $^{16}$O. The $0d_{5/2}$ and $1s_{1/2}$ orbitals are close in binding energy in the vicinity of $^{16}$O, but their spacing increases substantially in lighter nuclei, with the 5/2$^+$ state moving more rapidly in excitation energy than the 1/2$^+$ state. The 1/2$^-$ state also moves rapidly with respect to the 1/2$^+$ state. This is illustrated for a subset of the experimental information (nuclei with seven neutrons) in Fig.~\ref{fig1}. In the present communication we focus on the behavior of the 1/2$^+$ and 5/2$^+$ single-neutron excitations.

We show, through an examination of the simple geometrical effects of finite binding, that it is the qualitatively different behavior of neutron $s$ states near threshold which plays an important role in determining the sequence of levels in loosely bound light nuclei. To explore the changing pattern of states, we examine the available data where there is only one neutron in the $1s0d$-shell, spanning a range of neutron binding energies. 

%----------------------------FIGURE 1---------------------------------
%----------------------------------------------------------------------------

The behavior of $s$ states near threshold has been commented on before. Bohr and Mottelson~\cite{bohrandmot} note ``{\it The orbits with small angular momentum and small binding energy spend an appreciable amount of time outside the nucleus and thus benefit less from an increase in the size of the potential than do the weakly bound orbits with large $\ell$}''. The effect of phase space near threshold was discussed by Wigner~\cite{wigner} in the context of its influence on other open channels (Wigner cusps). The probability for the occurrence of states near threshold was discussed by Baz~\cite{baz} and by Inglis~\cite{inglis} in the context of the first narrow state occurring in $^5$He at 16.84 MeV, just above the D+T threshold. In 1964, Barker~\cite{barker2} had mentioned it in the context of the isospin-dependence of $s$ states in light nuclei. More recently, the behavior of $s$ states near threshold was discussed by Ozawa~\cite{ozawa} in terms of a possible shell gap at $N=16$, and by Hamamoto and Mottelson~\cite{hamamoto1} in the context of various nuclear structure properties e.g., changing shell structure, pairing, deformation, and halos. Also related to the present work is a study by Sagawa {\it et al}.~\cite{esbensen} on the behavior of even-parity states in $N=7$ nuclei. They show that the curvature of the 1/2$^+$ state is evident in Hartree-Fock calculations, however, the discussion in their paper is in terms of pairing blocking and coupling to quadrupole core excitations. None of these discussions make mention of how successfully finite binding effects provide an explanation for the pattern of behavior observed in the data. 

The intrusion of the $1s_{1/2}$ orbital into the $0p$-shell was already noted by Talmi and Unna~\cite{talmi} 50 years ago, who interpreted it in terms of the residual effective interactions between nucleons. In the sense that the residual interaction includes any modifications from the free nucleon-nucleon interaction, it may indeed also approximate the effects of finite binding, albeit with some difficulty because of the non-linearity of the effect.  

The shell model has been remarkably successful in describing the behavior of stable and exotic nuclei in terms of effective forces that mock up the residual $NN$ interaction. The explicit recognition of the tensor component of the free $np$ force and its monopole component has provided a successful explanation for many of the changes in magic numbers in heavier nuclei~\cite{otsuka}. However, based on the accumulated data presented in this work, the magnitude of the observed effect seems considerably larger than can be accounted for by the tensor force.

\begin{figure}
\centering
\includegraphics[scale=0.64]{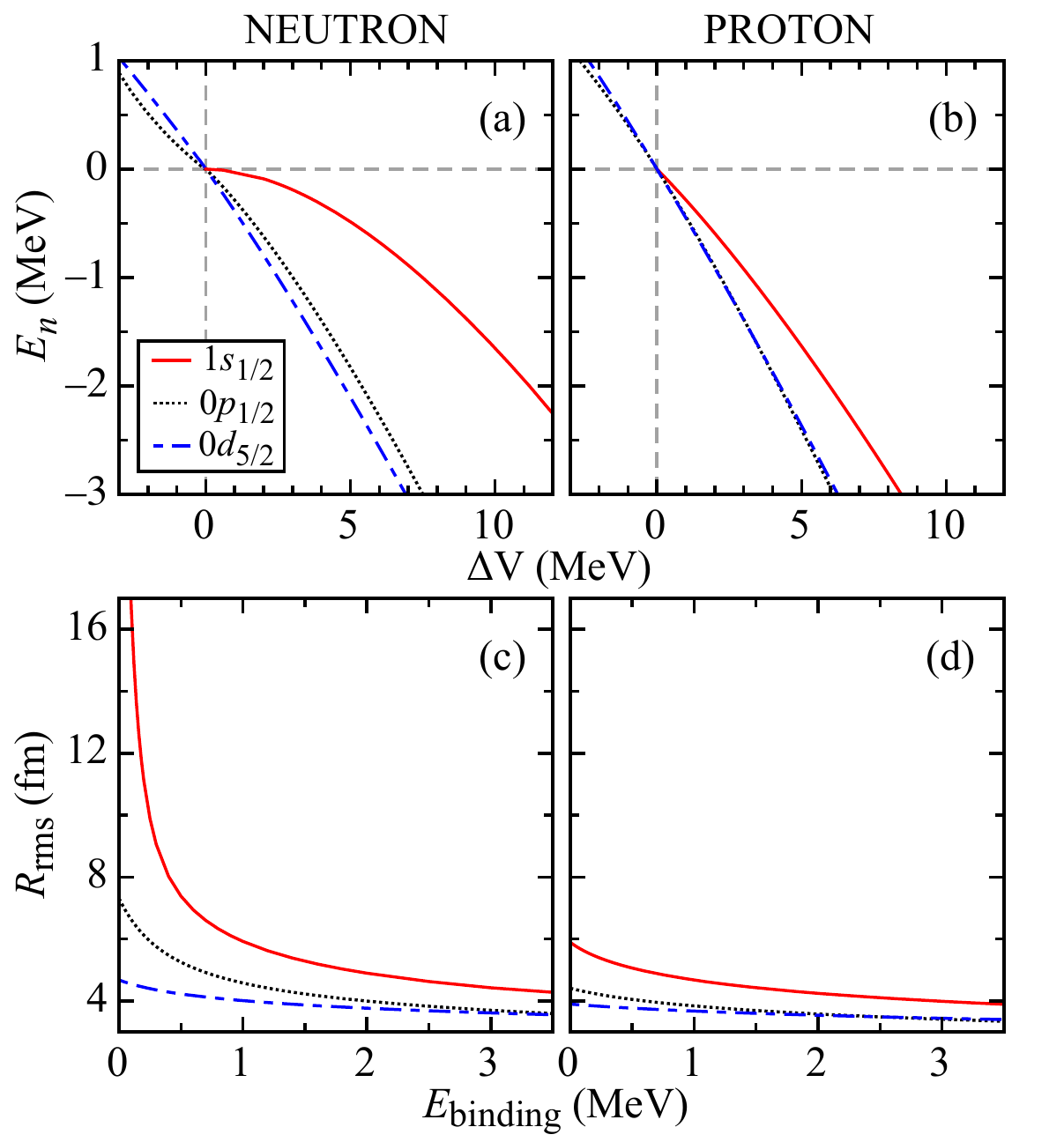}
\caption{\label{fig2}(color online). The calculated energies in Woods-Saxon potentials for of the $0p_{1/2}$, $1s_{1/2}$, and $0d_{5/2}$ single-nucleon excitations, $E_{n}$, as a function of $\Delta V$ are shown in panels (a) and (b) for neutrons and protons, respectively. $\Delta V$ is the change in potential depth required to move the energy of a state from $E_{n}=0$. Panels (c) and (d) show the corresponding rms radii as a function of binding energy.}
\end{figure}

To examine the behavior of single-nucleon states we use a Woods-Saxon potential shape with reasonable radial parameters (see Table~\ref{tab1}). In Fig.~\ref{fig2} we plot the calculated energy of the $0p_{1/2}$, $1s_{1/2}$, and $0d_{5/2}$ orbitals as a function of $\Delta V$, where $\Delta V \equiv V-V_0$ with $V_0$ being the potential strength needed to put the energy of a specific orbital at threshold. The behavior of the neutron 1/2$^+$ state near zero binding is quite dramatic:\ the same increment in the potential depth that increases the binding of the neutron 1/2$^-$ and 5/2$^+$ states from 0 to 1~MeV causes a change in the $s$ state that is almost an order of magnitude less. This influence of the threshold is by far the most pronounced for the neutron $s$ state---it plays a much smaller role for $\ell>0$ neutrons or for protons. The effect goes hand-in-hand with an increase in rms radius of the neutron density distribution, in other words, diffuse halo states~\cite{tanihata} are by far the most prominent for $\ell=0$ neutrons. In Fig.~\ref{fig2}, the large increase in rms radius for neutron $s$ states near threshold is evident, compared to those for protons or neutrons with finite $\ell$ values. 

The Woods-Saxon results suggest that this tendency of neutron $s$ states to linger just below threshold is a general property for any neutral particle with zero angular momentum, though the authors are not aware of any cases where this has been noted in other fields. 
 
 %----------------------------TABLE 1----------------------------------
\begin{table}
\caption{\label{tab1} Parameters used for defining the Woods-Saxon potential.}
\newcommand\T{\rule{0pt}{3ex}}
\newcommand \B{\rule[-1.2ex]{0pt}{0pt}}
\begin{ruledtabular}
\begin{tabular}{ccc}
$r_0$ (fm)\B & $a$ (fm) & $V_{\rm so}$ (MeV) \\
\hline
1.17 & 0.50 & 2.35 \\
1.25\footnote{Parameters in this row were used for the calculations in Fig.~\ref{fig2}; those in other rows to estimate uncertainties.} & 0.63 & 4.03 \\
1.30 & 0.80 & 6.11 \\
\end{tabular}
\end{ruledtabular}
\end{table}
%---------------------------------------------------------------------

%----------------------------FIGURE 2---------------------------------
%----------------------------------------------------------------------------

Since the work of Talmi and Unna~\cite{talmi} on the behavior of the 1/2$^+$ states in light nuclei, a great deal of experimental information has been accumulated on this single-particle excitation. The 5/2$^+$ state is chosen in the present work for comparison because the $0p$ shell is being filled in most of these nuclei. Fortune~\cite{fort} summarized the available data on these two states in 1995 in the context of the anomalous Coulomb displacement energy for  1/2$^+$ states~\cite{Coul}, which is also a reflection of the different radii for $s$ states. In recent work with the ($d$,$p$) reaction that specifically identifies the single-neutron component of the wave functions, and utilizing radioactive beams, additional information was obtained for $^9$He~\cite{9he}, $^{14}$B~\cite{b14}, and $^{16}$C~\cite{c16}.

%----------------------------FIGURE 3---------------------------------
\begin{figure}
\centering
\includegraphics[scale=0.65]{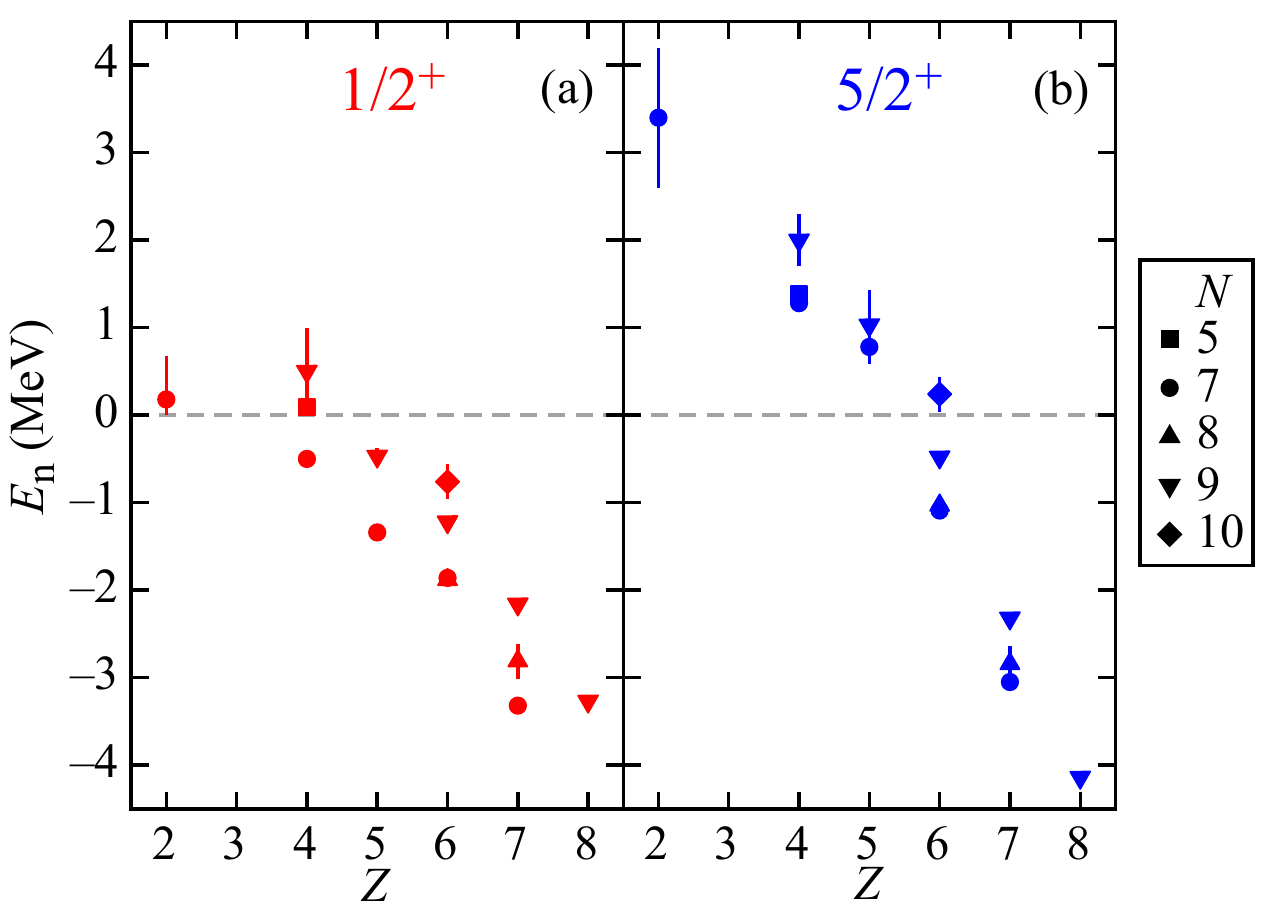}
\caption{\label{fig3}(color online). The available experimental data, $E_{n}$, on  the 1/2$^+$ (a) and 5/2$^+$ (b) states for isotopes of He to O with $N=5$--10 as a function of $Z$. The sources of the data are given in the Appendix.}
\end{figure}
%-----------------------------------------------------------------------------

This information, together with what was previously available provides data for 14 sets of single-particle energies and these are used in the present study. The data, together with the estimates of uncertainties, are given in the Appendix, and are presented in Fig.~\ref{fig3}. Further details are given in the Supplemental Material~\cite{supmat}. Data for $N=11$ and $N=13$ nuclei, where the $sd$ orbits are occupied by more than one neutron and therefore not directly comparable to the other data, are not included, though their behavior is similar.

Immediately apparent from Fig.~\ref{fig3} is the major trend in the neutron binding energy of these excitations and its dependence on the proton number $Z$. The stronger dependence of overall binding on proton number compared to neutron number $N$ is to be expected, as the monopole part of the total $np$ interaction is much stronger than the $nn$ part: the overall binding of the neutrons, the mean field, is influenced much more by the number of protons.

The 5/2$^+$ excitation energies range from being bound by $\sim$4~MeV to unbound by $\sim$3.5 MeV [Fig.~\ref{fig3}(b)].  The 1/2$^+$ excitation also becomes unbound in $^9$He, $^9$Be, and $^{13}$Be [Fig.~\ref{fig3}(a)]. (The energy for an unbound $\ell=0$ neutron single-particle resonance is difficult to define precisely from the available data.)

%----------------------------FIGURE 4---------------------------------
\begin{figure}[h]
\includegraphics[scale=0.63]{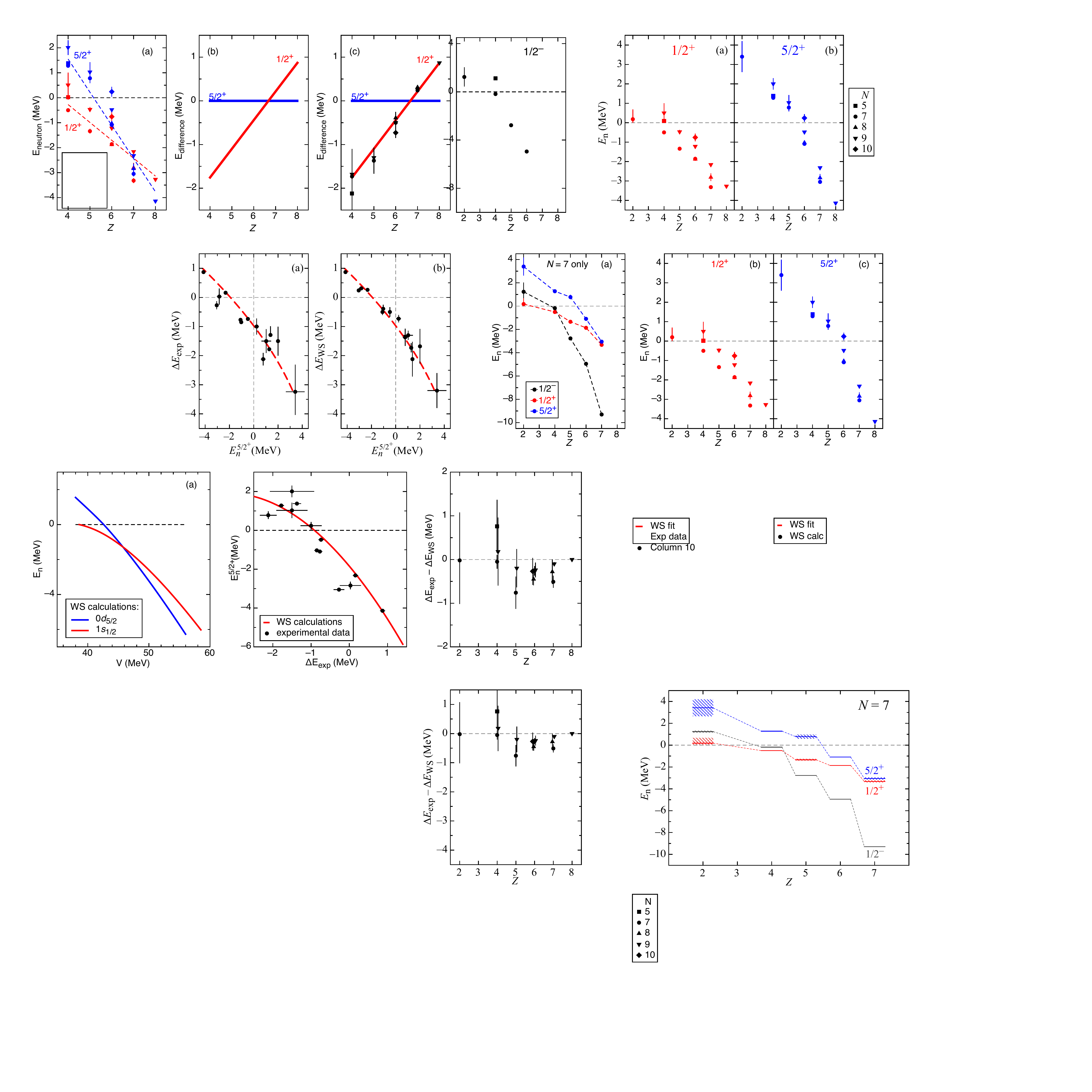}
\caption{\label{fig4}(color online). The difference between the experimentally determined 1/2$^+$ and 5/2$^+$ states, $\Delta E_{\rm exp}\equiv E_n^{1/2^+}-E_n^{5/2^+}$, as a function of the energy of the 5/2$^+$ (a), and the same difference calculated in a Woods-Saxon potential, $\Delta E_{\rm WS}$  (b). The dashed line is a smooth curve fit to the Woods-Saxon calculations and is the same in the two plots.}
\end{figure}
%-----------------------------------------------------------------------------

To explore the degree to which geometric effects play a role in the difference between the 1/2$^+$ and 5/2$^+$ excitations, we calculate the single-particle energies in a potential using a Woods-Saxon shape~\cite{ptolemy,pieper,volya} with the parameters in Table~\ref{tab1}.  We constrained the spin-orbit potential, with $r_{\rm so0}=1.10$~fm and $a_{\rm so}=0.50$~fm, to fit the levels of $^{17}$O and obtained $V_{\rm so}=4.03$~MeV (to the extent that $^{16}$O is a closed shell there is no tensor force effect here). The radius for each nucleus was taken as $r_0A^{1/3}$, and the central well depth was adjusted to fit the experimental 5/2$^+$ energies. Then the 1/2$^+$ energies were obtained in the same potential. The results are shown in Fig.~\ref{fig4}. Where the 5/2$^+$ excitation is unbound, the well was adjusted to yield a resonance in the elastic scattering at the appropriate energy in the continuum. A problem arises in $^{9}$He and in $^{13}$Be where the 1/2$^+$ excitation is not bound in the potential.  We therefore assumed, somewhat arbitrarily, that the state is just above threshold, similar to $^9$Be where the same assumption has been made historically, though with a quantitative justification that is not entirely clear. (We should note that deformation effects are not included, though they will play some role in the Be nuclei, with the largest change being the lowering of the 5/2$^+$ state.)

To get a rough estimate of the uncertainty in the energies obtained from the Woods-Saxon calculations, they were repeated with two other sets of radial parameters shown in rows 1 and 3 of Table~\ref{tab1}. For the cases where the 1/2$^+$ state was near threshold the calculated energy difference varied by as much  $\sim0.6$~MeV (in the cases of  $^9$He, $^9$Be, and $^{13}$Be) while it was much less for $S_n\gtrsim1$~MeV.   These variations were used to estimate the uncertainties in the calculated values of $\Delta E_{\rm WS}$ shown in Fig.~\ref{fig4}(b).  

It is evident from a comparison of Fig.~\ref{fig4}(a) and \ref{fig4}(b) that much of the variation in $\Delta E$ arises from finite binding effects. This is the dominant mechanism determining the ordering of the $1s_{1/2}$ and $0d_{5/2}$ orbitals in these nuclei. The similarity in the behavior of the calculated $0d_{5/2}$ and $0p_{1/2}$ orbits in Fig.~\ref{fig2}(a) implies that a major part of the intrusion of the 1/2$^+$ state into the $0p$ shell also arises from these finite binding effects. However, the tensor force complicates the pattern somewhat because of the large overlap between $p$-shell protons and the neutron $0p_{1/2}$ orbital. 

%----------------------------FIGURE 5---------------------------------
\begin{figure}[h]
\includegraphics[scale=0.78]{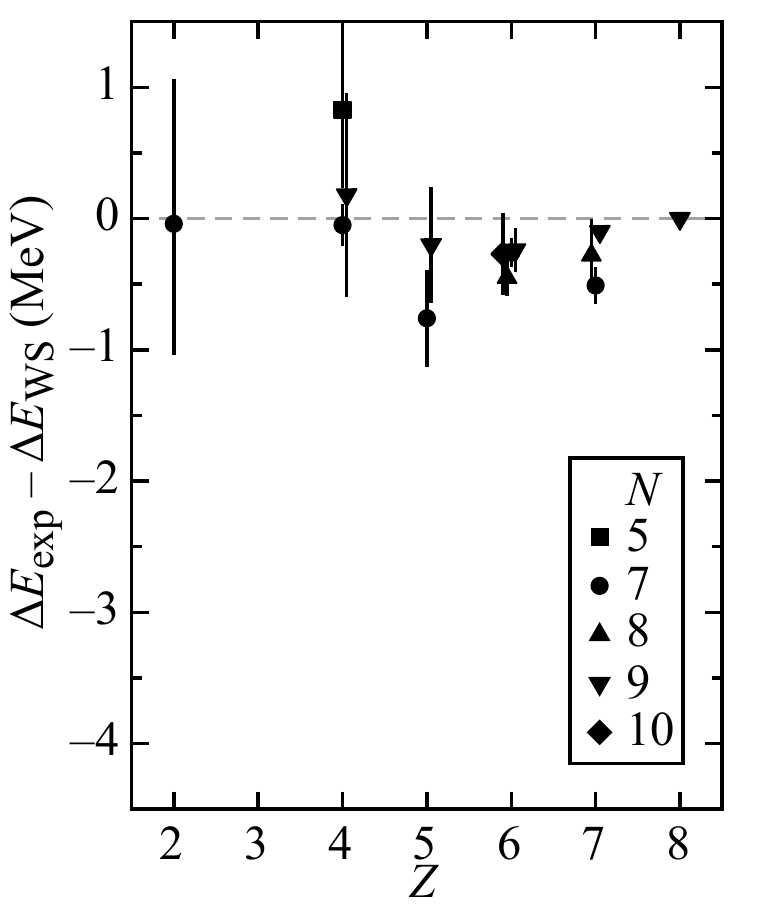}
\caption{\label{fig5}(color online). The difference between the $\Delta E_{\rm exp}$ and $\Delta E_{\rm WS}$ values from Fig.~\ref{fig4} plotted as a function of $Z$.} 
\end{figure}
%-----------------------------------------------------------------------------

In Fig.~\ref{fig5} we plot the difference between the actual 1/2$^+$-5/2$^+$ separations, $\Delta E_{\rm exp}$, and the Woods-Saxon ones, $\Delta E_{\rm WS}$, as a function of $Z$. While this difference is small, with relatively large uncertainties, it is of the right order of magnitude compared to what might be expected from the monopole component of the tensor force~\cite{otsuka} acting on the $0d_{5/2}$ excitation. Filling of the $0p_{3/2}$ shell between  $Z=2$ and 6 seems to result in a downward trend in Fig.~\ref{fig5} as expected from the tensor interaction, while the addition of $0p_{1/2}$ protons between $Z=6$ and 8 causes it to act in the opposite direction. For He and O one expects no effect from the $np$ tensor force, because of the closed oscillator shells of protons in these nuclei. While the spin-orbit strength was chosen to fit $^{17}$O, the difference is almost zero for  $^{9}$He, but the uncertainties are large.  

A closely related issue that has received considerable attention is that the large radii for $s$ states give rise to the anomalous Coulomb-energy differences known as the Thomas-Ehrman shift~\cite{Coul}. As mentioned above, Fortune~\cite{fort} surveyed the available data and carried out a Woods-Saxon calculation~\cite{fort1} to predict the low-lying mirror states of $^{11}$Be in $^{11}$N, much in the spirit of the present work.  Also related are some of the publications of Barker~\cite{barker} in this context and of Millener~\cite{millener}.

Oscillator shell-model calculations have, on the whole, been enormously successful~\cite{sm}. They are based on wave functions in an infinitely deep potential, with empirical effective interactions. These interactions are adjusted to fit data, and to the extent that the data are in the vicinity of threshold, they will reflect this. However, when there is a simple physical assumption that is not a part of the model, and is relevant to a prominent feature, it needs to be noted. In the case of these light nuclei the effects of finite binding have to be considered in describing a major change in shell structure.

The behavior of $s$ wave neutrons near threshold is the same physics responsible for the halo phenomenon, including neutron-pair halos such as in $^{11}$Li.  This is implicit in the discussions of Refs.~\cite{hamamoto1,tanihata}. The tendency for these $s$ states to linger near threshold also suggests that ground states with halos may occur over a relatively larger region of nuclei than might be estimated otherwise. Does a similar region occur near $N=50$, where the $2s_{1/2}$ neutron state may dive into the shell gap below the $1d_{5/2}$ and $0g_{7/2}$ states in neutron-rich nuclei? This may occur around $^{78}$Ni which is likely to become accessible to experiments soon. Or might it even occur with the $3s_{1/2}$ state beyond $N=126$ around $^{204}$Pt or $^{202}$Os?

The authors thank S.~C.~Pieper and A.~Volya for their help with the Woods-Saxon calculations, J.~Millener, N. Orr, and  A. Macchiavelli for calling our attention to relevant literature, and a number of other colleagues for helpful discussions. This work was supported by the US Department of Energy, Office of Nuclear Physics, under Contract No. DE-AC02-06CH11357.

\appendix*
\section{}

Information on the energy centroids for the $1/2^+$ and $5/2^+$ neutron excitations is given in Table~\ref{tab2}. The methods used to extract this information from the available data are given where necessary. For the $1/2^-$ excitations, only the origins of the levels shown in Fig.~\ref{fig1} are discussed. Further details on the $1/2^+$ and $5/2^+$ neutron excitations are given in the Supplemental Material~\cite{supmat}. 

%----------------------------TABLE2---------------------------------
\begingroup
\squeezetable
\begin{table*}
\caption{\label{tab2} Numerical quantities used in this study. All quantities are in MeV and only those with uncertainties that are 50~keV or greater are listed.}
\vspace*{1.5mm}
\newcommand\T{\rule{0pt}{3ex}}
\newcommand \B{\rule[-1.2ex]{0pt}{0pt}}
\renewcommand{\tabcolsep}{0.02cm}
\begin{ruledtabular}
\begin{tabular}{lcccccccc}
$^{A}Z\B$ & $S_n$ & $E_x^{1/2^+}$ & $E_x^{5/2^+}$ & $E_n^{1/2^+}$ & $E_n^{5/2^+}$ & $\Delta E_{\rm exp}$ & $\Delta E_{\rm WS}$ & Ref(s). \\
\hline
$^{9}$He\T & $-0.18^{+0.18}_{-0.50}$ & 0
 & $3.24^{+0.93}_{-0.80}$ & $0.18^{+0.50}_{-0.18}$ & $3.42\pm0.78$ & $-3.24^{+0.93}_{-0.80}$ & $-3.20\pm0.60$ & \cite{9he} \\

$^{9}$Be\T\B & 1.66 & $1.75^{+0.30}_{-0.05}$ & 3.05 & $0.09^{+0.30}_{-0.05}$ & 1.38 & $-1.29^{+0.30}_{-0.05}$ & $-2.12\pm0.60$ & \cite{Til04,Bur10} \\

$^{11}$Be& 0.50 & 0 & 1.78 & $-0.50$ & 1.28 & $-1.78$ & $-1.73\pm0.16$ & \cite{Aut70,Kel12} \\

$^{13}$Be & $-0.50\pm0.50$ & 0 & $1.50\pm0.50$ & $0.50\pm0.50$ & 2.00 & $-1.50\pm0.50$ & $-1.68\pm0.60$ & \cite{For13,Aks13,Kor95} \\

$^{12}$B &  3.37 & $2.03\pm0.10$  & $4.15\pm0.20$ & $-1.34\pm0.10$ & $0.78\pm0.20$ & $-2.12\pm0.22$ & $-1.37\pm0.30$ & \cite{Ajz90,Mon71,Lee10} \\
$^{14}$B &  0.97 & $0.50\pm0.10$  & $2.00\pm0.40$ & $-0.47\pm0.10$ & $1.03\pm0.40$ & $-1.50\pm0.41$ & $-1.30\pm0.16$ & \cite{b14,Kan05} \\
$^{13}$C &  4.95 & 3.09  & 3.85  & $-1.86$ & $-1.09$ & $-0.77$ & $-0.50\pm0.11$ & \cite{Ajz91} \\
$^{14}$C &  8.18 & 6.29  & 7.16  & $-1.88$ & $-1.03$ & $-0.85$ & $-0.39\pm0.14$ & \cite{Ajz91,Ajz86} \\
$^{15}$C &  1.22 & 0  & 0.74  & $-1.22$ & $-0.48$ & $-0.74$ & $-0.50\pm0.17$ &\cite{Ajz91}  \\
$^{16}$C &  4.25 & $3.49\pm0.20$  & $4.49\pm0.20$ & $-0.76\pm0.20$ & $0.24\pm0.20$ & $-1.00\pm0.28$ & $-0.73\pm0.13$ & \cite{c16,Til93,Li76} \\
$^{14}$N  &  10.55 & $7.23\pm0.10$  & $7.50\pm0.10$ & $-3.32\pm0.10$ & $-3.05\pm0.10$ & $-0.27\pm0.14$ & 0.24 & \cite{Ajz91}\\
$^{15}$N  &  10.83 & $8.02\pm0.20$  & $7.99\pm0.20$ & $-2.81\pm0.20$ & $-2.84\pm0.20$ & $0.03\pm0.28$ & 0.31 & \cite{Ajz91,Phi69,Mer13} \\
$^{16}$N  &  2.49 & 0.33  & 0.17  & $-2.16$ & $-2.32$ & 0.16  & 0.26 & \cite{Til93,Boh72,Bar08} \\
$^{17}$O  &  4.14 & 0.87  & 0  & $-3.27$ & $-4.14$ & 0.87  & 0.87 & \cite{Til93,Ajz77}\\
\end{tabular}
\end{ruledtabular}
\end{table*}
\endgroup
%-----------------------------------------------------------------------------

The neutron separation energies were all taken from Ref.~\cite{Aud12} with the exception of $^{9}$He, where it was taken to be $-0.18$~MeV~\cite{9he}.

In the majority of cases the 1/2$^+$, 5/2$^+$, and 1/2$^-$ excitations have been observed in transfer reactions and carry essentially all the single-particle strength and thus those states are taken as the centroid energy of the respective excitation. Where several states carry the single-particle strength, the $(2J+1)C^2S$-weighted average of the corresponding spin-multiplet was taken from the references given in Table~\ref{tab2}.

In several of the nuclei used in this study the 1/2$^+$, 5/2$^+$, and 1/2$^-$ neutron excitations are unbound and require discussion because of possible ambiguities. In the case of $^{9}$He, the 1/2$^+$ strength has been attributed to a low-lying resonance at $E_n=180\pm85$~keV~\cite{9he}. We somewhat arbitrarily take 0.18$^{+0.50}_{-0.18}$~MeV as the resonance energy. The 1/2$^-$ resonance is at 1.24(10) MeV and the width is consistent with the Wigner limit~\cite{9he}.  The 1/2$^+$ excitation in $^{9}$Be at 1.66~MeV~\cite{Aud12} had been accepted as the neutron single-particle state for some time. The 1/2$^+$ excitation in $^{13}$Be is taken as the ground state and has been attributed to a low-lying resonance at an energy of $E_n\approx0.50\pm0.50$~MeV~\cite{For13,Aks13}. The widths of the unbound 5/2$^+$ states $^{9}$He and $^{9,11,13}$Be are generally consistent with the single-particle Wigner limit.

For two of the nuclei surveyed, the extraction of the neutron excitations was not trivial. $^{16}$C has two neutrons in the $sd$ shell, but by combining information with the experimental data on $^{17,18}$O the residual interaction between the two neutrons can be removed, and thus the numbers are directly comparable with cases where there was only a single neutron in this shell. These are the only data included with $N=10$. We make use of the fact that the low-lying 3$^+$ state in $^{16}$C $must$ be of the $sd$ configuration, while the 4$^+$ must be $d^2$.  The 1/2$^+$ energy was determined from the 3$^+$ state in $^{16}$C at 4.09~MeV~\cite{c16,Til93}, and a correction of 0.60~MeV was made using the $^{18}$O $(0d_{5/2}1s_{1/2})_{J=3}$ matrix element~\cite{Li76}. We estimate an uncertainty of 200~keV. The 5/2$^+$ energy was determined from the 4$^+$ state in $^{16}$C at 4.14~MeV~\cite{Til93}, after a correction of $-$0.35~MeV to the centroid was made using the $^{18}$O $(0d_{5/2})^2_{J=4}$ matrix element~\cite{Li76}. We estimate an uncertainty of 200~keV.

The other non-trivial case is $^{14}$N, the only nucleus included in which $N=Z$. We need the energies of the neutron excitation with respect to the $^{13}$N (spin 1/2$^-$) core, which means the centroids from the 0$^-$ and 1$^-$ states for the $s_{1/2}$ excitation and that from the 2$^-$ and 3$^-$ states for the $d_{5/2}$.  In {$^{14}$N} these are split by isospin into $T=0$ and  $T=1$ components.  With good isospin, the configurations are mixtures of the configuration where the even-parity nucleon can be either a neutron or a proton. The neutron energies are given by $E=1/2E_{T=0}+1/2E_{T=1}$ but since the Coulomb force breaks the isospin symmetry a correction needs to be applied for the $s-d$ difference in Coulomb energies for the component that involves protons. Since the relevant $T=1$ states in $^{14}$N are not well determined, we make use of the states in $^{14}$C. The 1/2$^+$ energy centroid is the $(2J+1)C^2S$-weighted average (with $C^2S=1$ assumed in this case) of the $T=0$, 0$^-$ and $1^-$ states at 4.92 and 5.69~MeV, identified in the $^{13}$C($^3$He,$d$)$^{14}$N reaction~\cite{Ajz91} and the $T=1$, 1$^-$ and 0$^-$ states at 8.06 and 8.78~MeV, identified as analogs to the states populated in the $^{13}$C($d$,$p$)$^{14}$C reaction~\cite{Ajz91}. A Coulomb-energy correction of 0.36~MeV has to be applied; this is half the difference between the 5/2$^+$ excitation energies in the $^{13}$C--$^{13}$N mirror nuclei~\cite{Ajz91}. Because there is evidence for some fragmentation of transfer strength a 100~keV uncertainty is assigned. The 5/2$^+$ energy centroid is that of the $T=0$, 2$^-$ and $3^-$ states at 5.11 and 5.83~MeV, identified in the $^{13}$C($^3$He,$d$)$^{14}$N reaction, the $T=1$, 3$^-$ and 2$^-$ states at 8.91 and 9.51~MeV, identified as analogs to the states in  $^{13}$C($d$,$p$)$^{14}$C with all the data from~\cite{Ajz91}. A Coulomb energy correction of 0.15~MeV has to be applied; half the difference between the 1/2$^+$ excitation energies in the $^{13}$C--$^{13}$N mirror nuclei. Because there is evidence for some fragmentation of transfer strength a 100 keV uncertainty is assigned. Since this involves two isospin states of the $(p_{1/2})^2$ configuration the 1/2$^-$ energy centroid is the mean of the 1$^+$ ground state and 0$^+$ first excited state, with no weighting by spin or isospin.

\end{document}